\documentclass[aps,prd,twocolumn,groupedaddress]{revtex4-1}

\usepackage{amssymb}
\usepackage{amsmath}
\usepackage{graphicx}
\usepackage{enumerate}
\usepackage{mathtools}
\usepackage{subfigure}
\usepackage{bm}

\begin{document}

\title{Derivation of the sterile neutrino Boltzmann equation from quantum kinetics}

\author{Lucas Johns}
\email[]{ljohns@physics.ucsd.edu}
\affiliation{Department of Physics, University of California, San Diego, La Jolla, California 92093, USA}

%\date{\today}

\begin{abstract}
An extensive, growing body of work has been penned on cosmologies that include one or more sterile neutrinos. Early entries in the literature formulated a Boltzmann-like equation describing sterile-neutrino production in a way that bypasses the numerical tracking of high-frequency complex phases, and meticulous quantum-kinetic analyses shortly thereafter put the formula on firmer ground. A new and more direct derivation is given here, showing that the equation follows almost immediately from a quantum relaxation-time approximation and an expansion in the mixing angle. Besides reproducing the desired result, the relaxation ansatz captures to a high degree of accuracy the interlaced dynamics of oscillations, decoherence, and plasma repopulation.  Successes and limitations of the semiclassical equation are illustrated numerically and are shown to reflect the accuracy of the approximations employed in the derivation. The inclusion of interactions among the sterile neutrinos is also briefly addressed.
\end{abstract}

\maketitle

\section{Introduction}

Sterile neutrinos continue to be actively studied as sources of oscillation anomalies, as reconcilers of cosmic tensions, and as candidates for dark matter. In all these cases the cosmological abundance must be calculated, and so the dynamics of active--sterile mixing must be contended with. The essential challenge is that the full quantum-kinetic problem involves disparate time scales and the interplay of coherent ($\propto G_F$) and incoherent ($\propto G_F^2$) effects.

A Boltzmann-like equation is often used to calculate the nonthermal abundance of sterile neutrinos produced from active ones \cite{kainulainen1990, cline1992, dodelson1994}:
\begin{equation}
\frac{d f_s}{dt} = \frac{\Gamma_a}{4} \frac{\sin^2 2\theta_m}{1 + \left( \frac{\Gamma_a}{2 \omega_m} \right)^2} \left( f_a - f_s \right), \label{boltz1}
\end{equation}
where $f_{a(s)}$ is the active (sterile) distribution function, $\Gamma_a$ is the scattering rate of active neutrinos, and $\theta_m$ and $\omega_m$ are the in-medium mixing angle and oscillation frequency. (We suppress dependence on momentum here and throughout.) Eq.~\eqref{boltz1} is a semiclassical approximation of the quantum kinetic equation (QKE) \cite{stodolsky1987, mckellar1994, enqvist1990, enqvist1991, raffelt1993, sigl1993, strack2005, boyanovsky2007, vlasenko2014}
\begin{equation}
i \frac{d \rho}{dt} = \left[ H , \rho \right] + i C \label{qke1}
\end{equation}
for the density matrix $\rho$. Its computational appeal lies in the fact that, by packaging together the effects of the Hamiltonian $H$ and collision term $C$ as a single effective production rate, one can overlook the quantum phases and evolve only the classical densities.

The first derivations of Eq.~\eqref{boltz1} (or variations of it) were based on single-particle arguments that equated the $\nu_s$ production rate with the product of the $\nu_a$ scattering rate and the probability of a $\nu_a$ oscillating into a $\nu_s$ \cite{kainulainen1990, cline1992}. Later analyses hearteningly arrived at similar formulas working from quantum-kinetic descriptions and judiciously applying approximations for the evolution in flavor space \cite{shi1996, foot1996, foot1997, bell1999, volkas2000, dibari2000, lee2000, dolgov2000}. Our purpose here is to add another entry to the list, one that is complementary to the references just cited and whose virtue is the insight it gives into the quantum dynamics underlying the semiclassical behavior. Given the ongoing interest in sterile neutrinos, having a robust simplification of the quantum dynamics may prove useful for future applications. The guiding idea, which we support numerically, is that the evolution of $\rho$ at small mixing angle is well described by the exponential decay of its deviations from equilibrium. As we demonstrate below, this simple ansatz leads promptly to Eq.~\eqref{boltz1}.

In Sec.~\ref{derivation} we go through the derivation and discuss it in the context of other treatments. In Sec.~\ref{numerics} we present numerical comparisons of the Boltzmann and QKE solutions, highlighting the accuracy not only of Eq.~\eqref{boltz1} but also of the ansatz on which it is based. In Sec.~\ref{conclusion} we conclude.

\begin{figure*}
\centering
\begin{subfigure}{
\centering
\includegraphics[width=.47\textwidth]{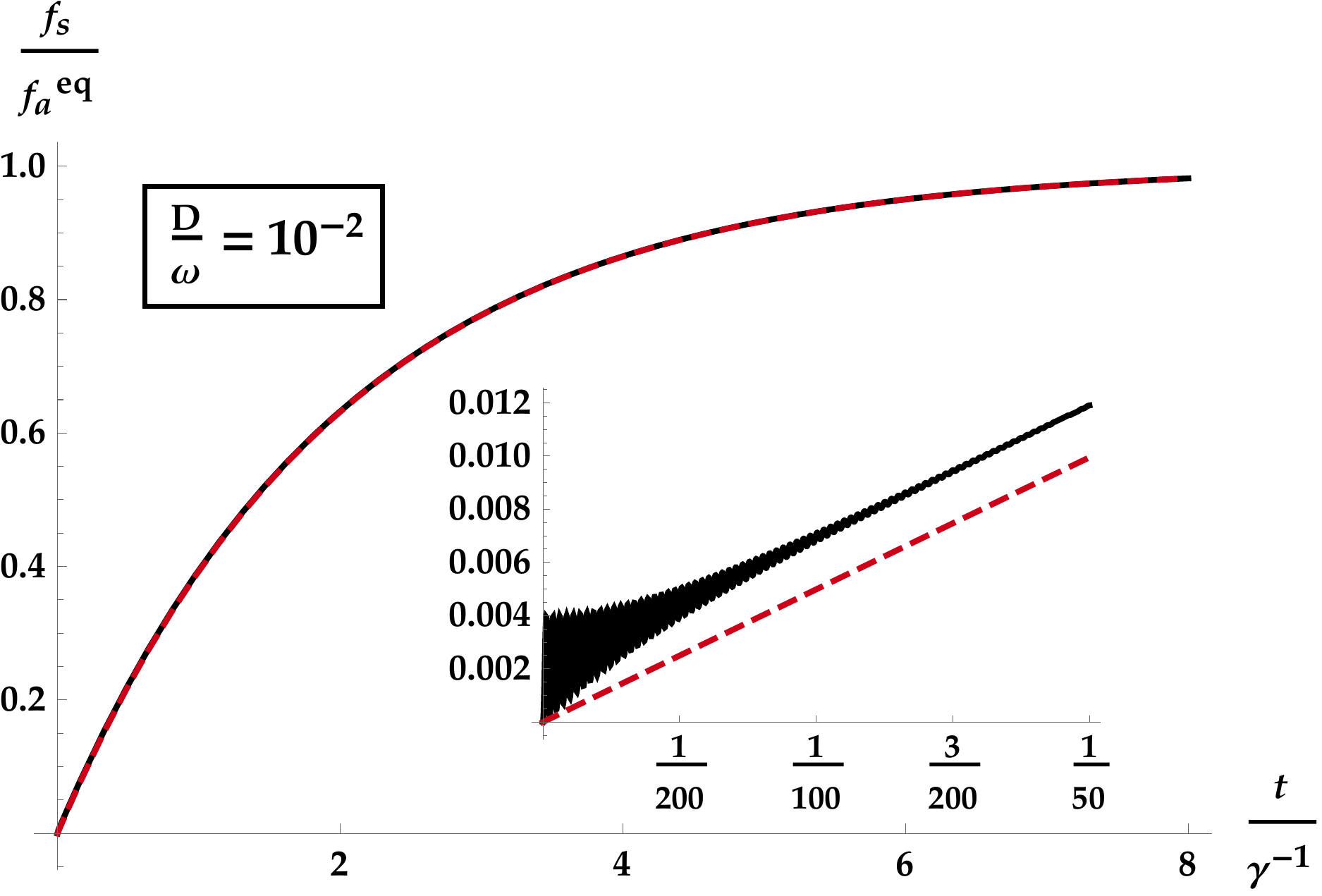}
}
\end{subfigure}
\begin{subfigure}{
\centering
\includegraphics[width=.47\textwidth]{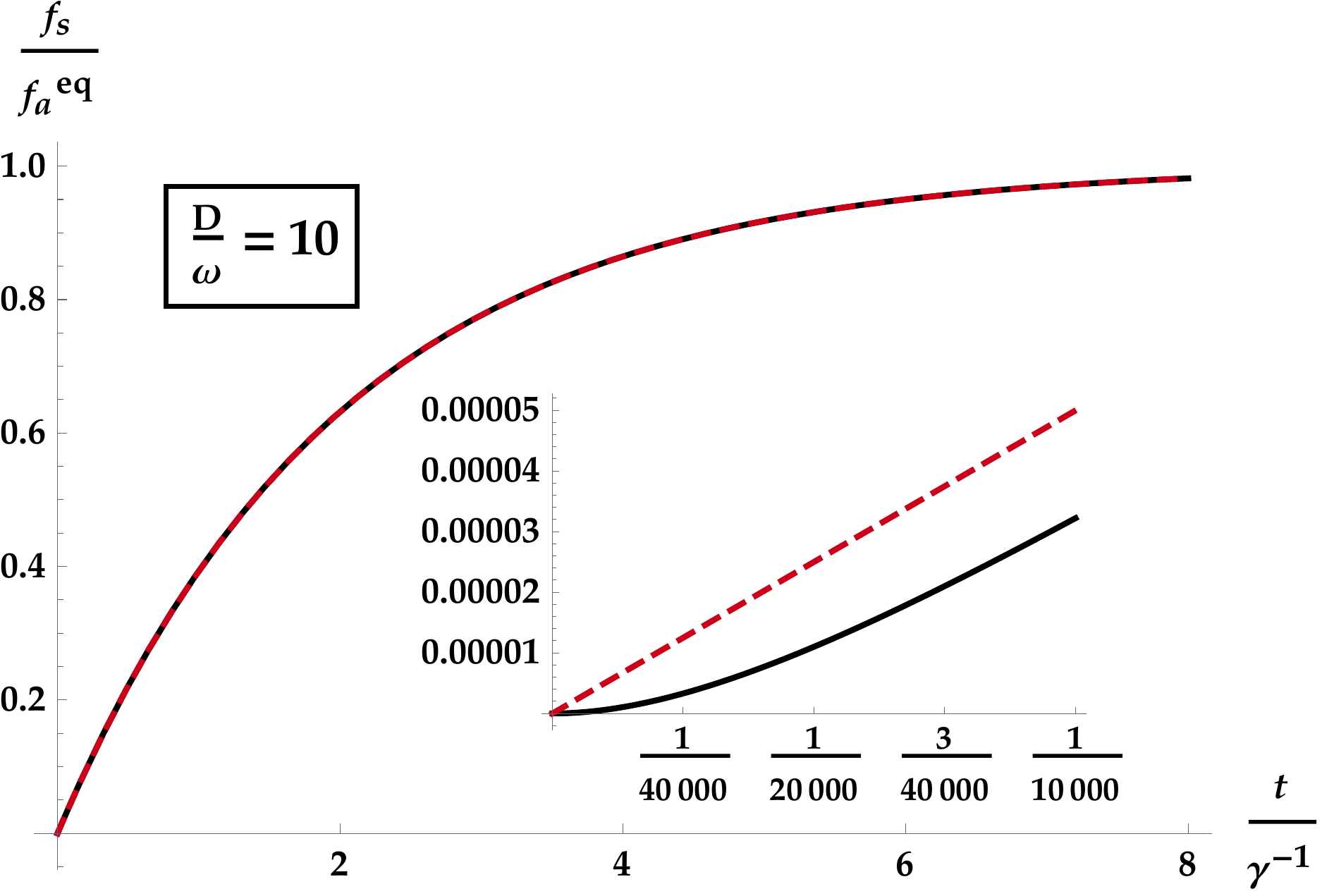}
}
\end{subfigure}
\caption{Comparison of the Boltzmann [Eq.~\eqref{boltz1}; dashed red curve] and QKE [Eqs.~\eqref{poleom}; black] solutions for $f_s (t)$. The latter solution uses the conversion $f_s = P_0 (1 - P_z) / 2$. The mixing angle is small: $\theta = \pi / 100$. Insets show the same curves on shorter time scales. On $\gamma^{-1}$ time scales the curves are indistinguishable by eye.}
\label{fsplots}
\end{figure*}

\begin{figure*}
\centering
\begin{subfigure}{
\centering
\includegraphics[width=.47\textwidth]{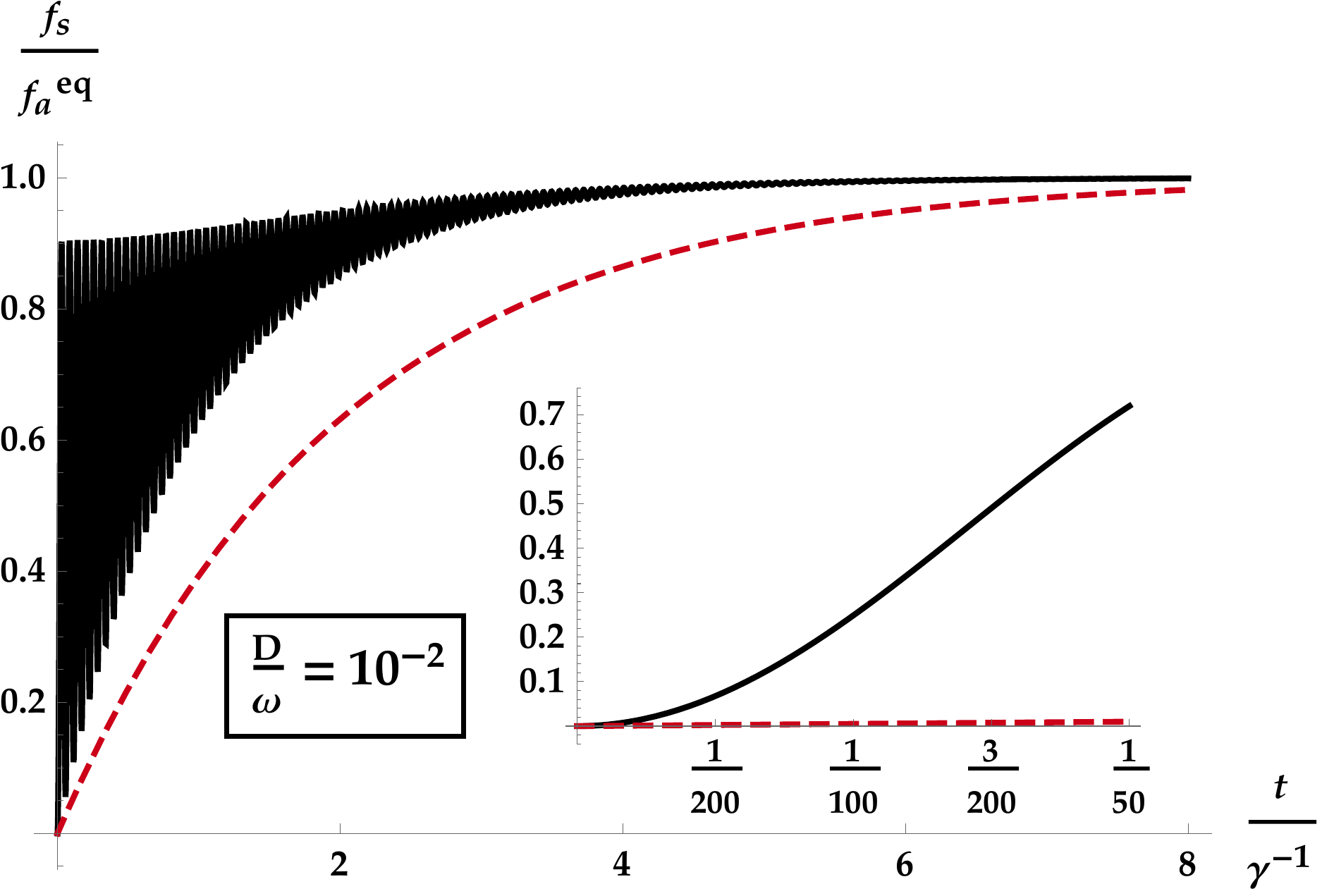}
}
\end{subfigure}
\begin{subfigure}{
\centering
\includegraphics[width=.47\textwidth]{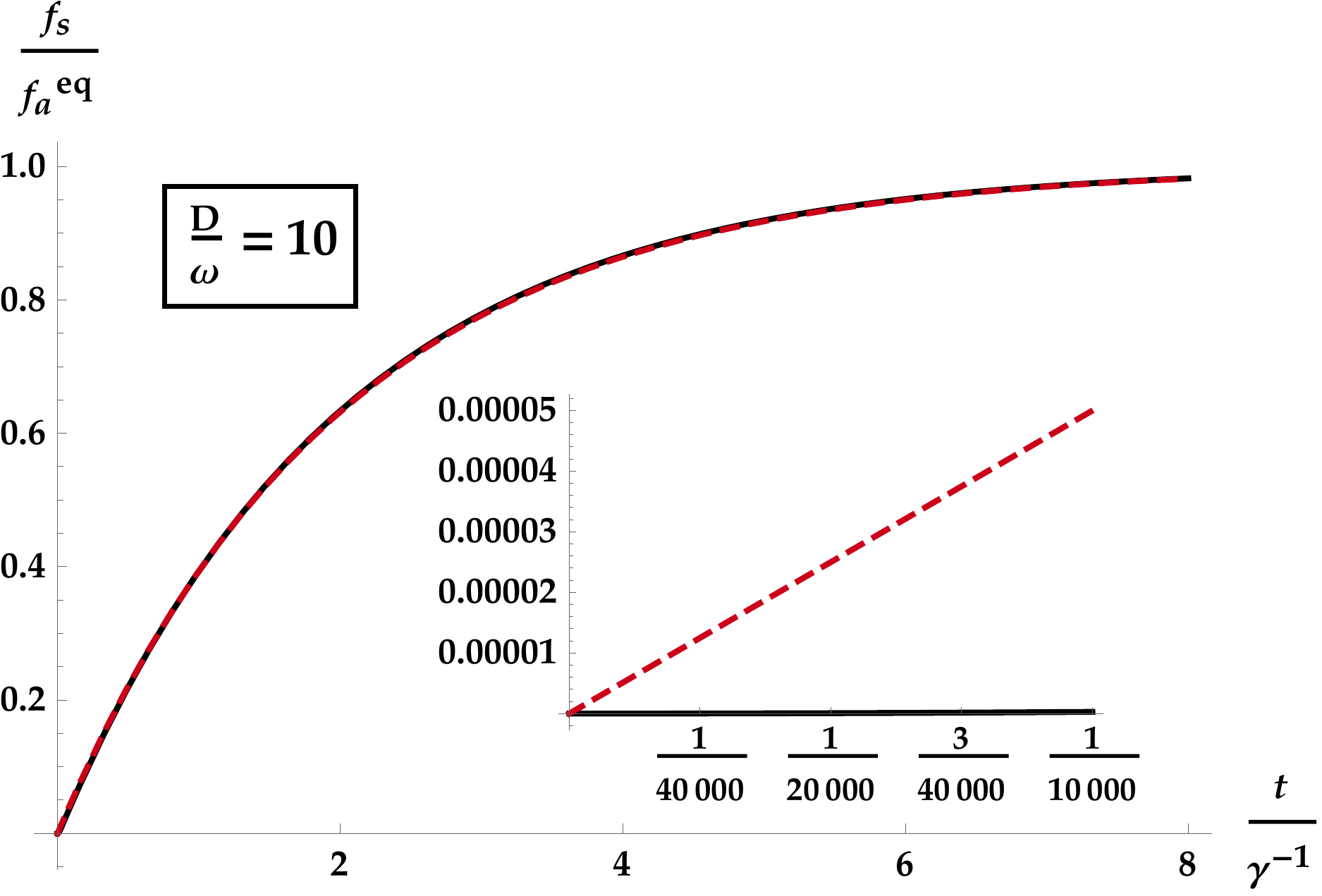}
}
\end{subfigure}
\caption{Same as Fig.~\ref{fsplots} but here the mixing angle is \textit{not} small: $\theta = \pi/5$. Unlike in the small-$\theta$ case, strong damping is necessary to coerce $\rho$ toward the Boltzmann solution. Early-time discrepancy is magnified compared to Fig.~\ref{fsplots}.}
\label{fslargeplots}
\end{figure*}

\section{Derivation \label{derivation}}

We begin by applying the quantum relaxation-time approximation to the collision term \cite{mckellar1994, bell1999, dolgov2001, hannestad2015}:
\begin{equation}
i C = \left\lbrace i \Gamma , \rho^\textrm{eq}_C - \rho \right\rbrace, \label{gammarelax}
\end{equation}
where $\Gamma = \left( 1 / 2 \right) \textrm{diag} ( \Gamma_a , \Gamma_s )$ and $\rho^\textrm{eq}_C$ is the $H = 0$ equilibrium. If the states do not mix, then the \textit{classical} relaxation-time approximation is recovered, and the densities $f_a$ and $f_s$ approach their equilibrium values with time scales $\Gamma_a^{-1}$ and $\Gamma_s^{-1}$ respectively. We posit that the same approximation applies to the entire right-hand side of Eq.~\eqref{qke1}, with a single effective relaxation parameter replacing the individual scattering rates and the flavor equilibrium $\rho^\textrm{eq}_F$ replacing the classical equilibrium. That is,
\begin{equation}
i \frac{d \rho}{dt} = \left\lbrace i \Gamma_\textrm{eff} , \rho^\textrm{eq}_F - \rho \right\rbrace,
\end{equation}
with $\Gamma_\textrm{eff} = \left( \gamma_m / 4 \right) \textrm{diag} \left( 1, 1 \right)$. (The extra factor of $1/2$ is added as a matter of preference.) Hence 
\begin{equation}
\frac{d \rho}{dt} = \frac{\gamma_m}{2} \left( \rho^\textrm{eq}_F - \rho \right), \label{relax}
\end{equation}
and in particular
\begin{equation}
\frac{d f_s}{dt} = \frac{\gamma_m}{2} \left( f_a^\textrm{eq} - f_s \right). \label{dfsdt}
\end{equation}
If the mixing angle is small, $f_a^\textrm{eq}$ can safely be replaced in this equation by $f_a$.

Using $\rho = P_0 \left( 1 + \mathbf{P} \cdot \boldsymbol{\sigma} \right) / 2$, it follows from Eq.~\eqref{relax} that the polarization vector obeys
\begin{align}
&\frac{d P_0}{dt} = \frac{\gamma_m}{2} \left( 2 f_a^\textrm{eq} - P_0 \right), \notag \\
&\frac{d \mathbf{P}}{dt} = - \gamma_m \frac{f_a^\textrm{eq}}{P_0} \mathbf{P}. \label{gameom}
\end{align}
At the same time, using $H = \left(\omega_m / 2 \right) B_m$ and setting $\Gamma_s = 0$, Eqs.~\eqref{qke1} and \eqref{gammarelax} imply
\begin{align}
&\frac{d P_0}{dt} = 2 D \left( f_a^\textrm{eq} - P_0 \frac{1 + P_z}{2} \right) \notag \\ 
&\frac{d \mathbf{P}}{dt} = \omega_m \mathbf{B}_m \times \mathbf{P} - D \mathbf{P}_T - \frac{\dot{P}_0}{P_0} \mathbf{P} + \frac{\dot{P}_0}{P_0} \mathbf{z}, \label{poleom}
\end{align}
where $D = \Gamma_a / 2$ is the decoherence rate and $\mathbf{B}_m = \sin 2\theta_m\mathbf{x} - \cos 2\theta_m\mathbf{z}$. The ansatz tells us that Eqs.~\eqref{gameom} and Eqs.~\eqref{poleom} can be set equal to each other at any moment in the evolution. For the sake of extracting $\gamma_m$, we choose to equate them prior to significant sterile production, during which time $\mathbf{P}$ nearly equals $\mathbf{z}$ and $P_0$ and $f_a$ nearly equal $f_a^\textrm{eq}$. To first order in the deviations, $\mathbf{P}$ satisfies the eigenvalue equation
\begin{equation}
\omega_m \mathbf{B}_m \times \mathbf{P} - D \mathbf{P}_T = - \gamma_m \mathbf{P}. \label{expeig}
\end{equation}
Nontrivial solutions of Eq.~\eqref{expeig} correspond to roots of the cubic equation
\begin{equation}
\gamma_m^3 - 2 D \gamma_m^2 + \left( D^2 + \omega_m^2 \right) \gamma_m - D \omega_m^2 \sin ^2 2\theta_m = 0.
\end{equation}
Applying perturbation theory to zeroth order in $\sin^2 2\theta_m$ uncovers two of the roots, with values $D \pm i \omega_m$. The third root, which is the purely real one that we seek, appears at first order:
\begin{equation}
\gamma_m = \frac{D \omega_m^2 \sin^2 2\theta_m}{\omega_m^2 + D^2}
\end{equation}
Plugging this into Eq.~\eqref{dfsdt}, we arrive at Eq.~\eqref{boltz1} as desired.

The analysis applies just as well to antineutrinos (or the right-handed states, if Majorana) as it does to neutrinos (or the left-handed states). If chemical potentials are involved in the dynamics, they are simply incorporated into the equilibrium distribution functions.

Eq.~\eqref{expeig} was also considered in Ref.~\cite{stodolsky1987} (albeit not in the context of sterile neutrinos), Ref.~\cite{dolgov2000} (albeit in a somewhat different form), and Refs.~\cite{shi1996, bell1999, volkas2000}. The last two were part of a series, along with Refs.~\cite{foot1996, foot1997, dibari2000, lee2000}, that provided significant insights into the dynamics of active--sterile oscillations. Vital to the derivation developed in those works is the approximation $d P_0 / dt = 0$, which was carefully shown in Ref.~\cite{lee2000} to be justified despite its inconsistency with $f_a$ remaining near equilibrium during sterile production. We similarly find that the correct value of $\gamma_m$ is obtained by dropping the repopulation terms from the equation of motion obeyed by $\mathbf{P}$, even though repopulation is crucial for accurately describing the evolution of the system as a whole. Our findings, based on the quantum relaxation-time approximation, are consistent in this regard with those of Ref.~\cite{lee2000}, based (in the words of the authors) on the ``brute-force'' approach. 

Besides reproducing Eq.~\eqref{boltz1} with minimal effort, the preceding analysis also has the advantage of pointing to a deeper physical picture: the small-$\theta_m$ dynamics is dominated by a flavor-space trajectory in which nonequilibrium deviations $\rho^\textrm{eq}_F - \rho$ decay with (instantaneous) lifetime $2 / \gamma_m$. Moreover, as we show in the next section, this approach sacrifices little in the way of accuracy for what it gains in simplicity.

But before moving on to the numerical analysis, let us briefly comment on the generalization to scenarios in which sterile neutrinos remain inert under the Standard Model couplings but interact via new ones. Reinstating $\Gamma_s \neq 0$ in Eq.~\eqref{gammarelax} leads to 
\begin{widetext}
\begin{gather}
\frac{d P_0}{dt} = 2 D_a \left( f_a^\textrm{eq} - P_0 \frac{1 + P_z}{2} \right) + 2 D_s \left( f_s^\textrm{eq} - P_0 \frac{1 - P_z}{2} \right) \notag \\
\frac{d \mathbf{P}}{dt} = \omega_m \mathbf{B}_m \times \mathbf{P} - \left( D_a + D_s \right) \mathbf{P}_T - \frac{\dot{P}_0}{P_0} \mathbf{P} + \bigg[ 2 D_a \left( f_a^\textrm{eq} - P_0 \frac{1 + P_z}{2} \right) - 2 D_s \left( f_s^\textrm{eq} - P_0 \frac{1 - P_z}{2} \right) \bigg] \mathbf{z}, \label{spol}
\end{gather}
\end{widetext}
where $D_{a,s} = \Gamma_{a,s} / 2$ and $f_s^\textrm{eq}$ denotes the equilibrium that $f_s$ tends toward if the mixing is turned off. Because repopulation of $f_s$ drives $\mathbf{P}$ toward $-\mathbf{z}$, the net effect of repopulation on $\mathbf{P}$ no longer takes the second-order form $( \dot{P}_0 / P_0 ) (\mathbf{z} - \mathbf{P} )$.

However, suppose that $\Gamma_s$ is much faster than the active--sterile conversion rate. In that case $f_s$ is always very close to $f_s^\textrm{eq}$ on the conversion time scale, and the new repopulation terms can once again be dropped from the equations so long as $f_s$ is consistently interpreted as being at the sterile-sector equilibrium value. The same relaxation ansatz can then be used as before, again leading to Eq.~\eqref{expeig}. The only difference now is that the decoherence rate in the expression for $\gamma_m$ should be interpreted as the sum $D_a + D_s$, in agreement with Refs.~\cite{hannestad2014, archidiacono2015, johns2019}. Note that the distinction between $\rho^\textrm{eq}_C$ and $\rho^\textrm{eq}_F$ is generally important to make, but is rendered moot when $\Gamma_s = 0$.

\begin{figure*}{
\begin{subfigure}{
\centering
\includegraphics[width=.47\textwidth]{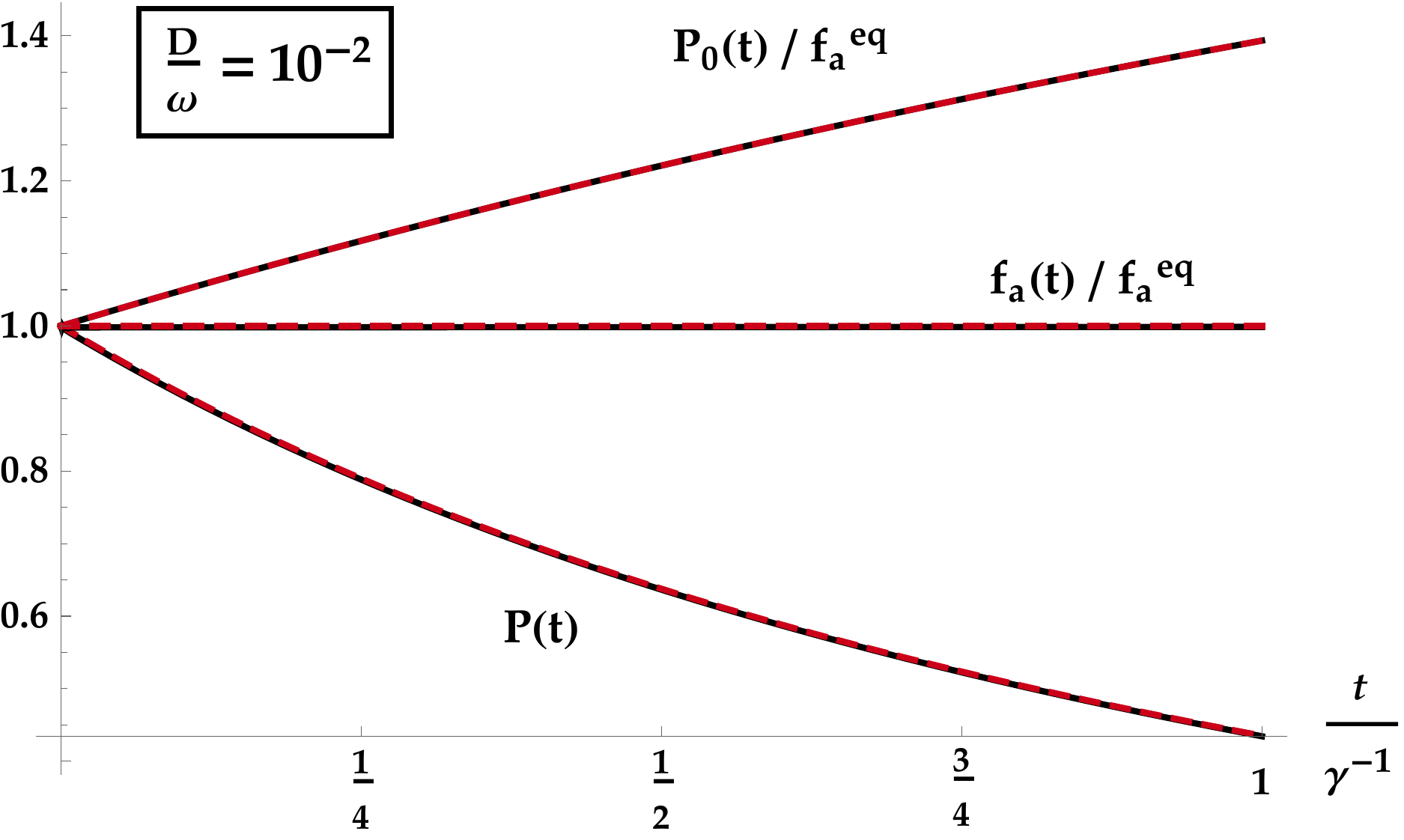}
}
\end{subfigure}
\begin{subfigure}{
\centering
\includegraphics[width=.47\textwidth]{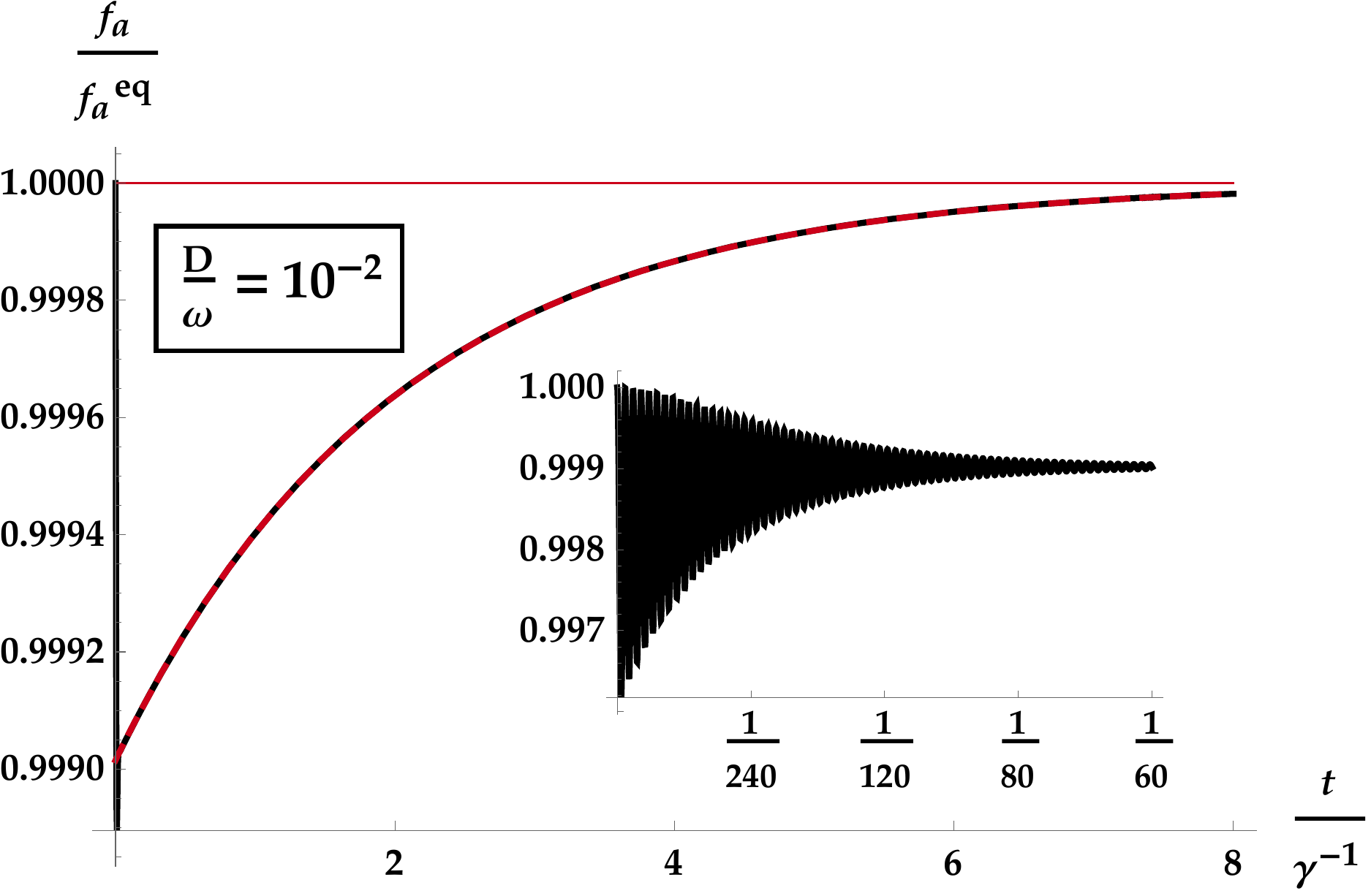}
}
\end{subfigure}
\caption{Comparison of the relaxation [Eq.~\eqref{relax}; dashed, red curve] and QKE [Eq.~\eqref{qke1}; black] solutions for $P_0 (t)$, $f_a (t)$, and $P (t) = | \mathbf{P} (t) |$. The thin red line in the right panel denotes the equilibrium value. The onset time of the relaxation solution is fit by hand.}
\label{magplot}
}
\end{figure*}

\begin{figure*}
\centering
\begin{subfigure}{
\centering
\includegraphics[width=.47\textwidth]{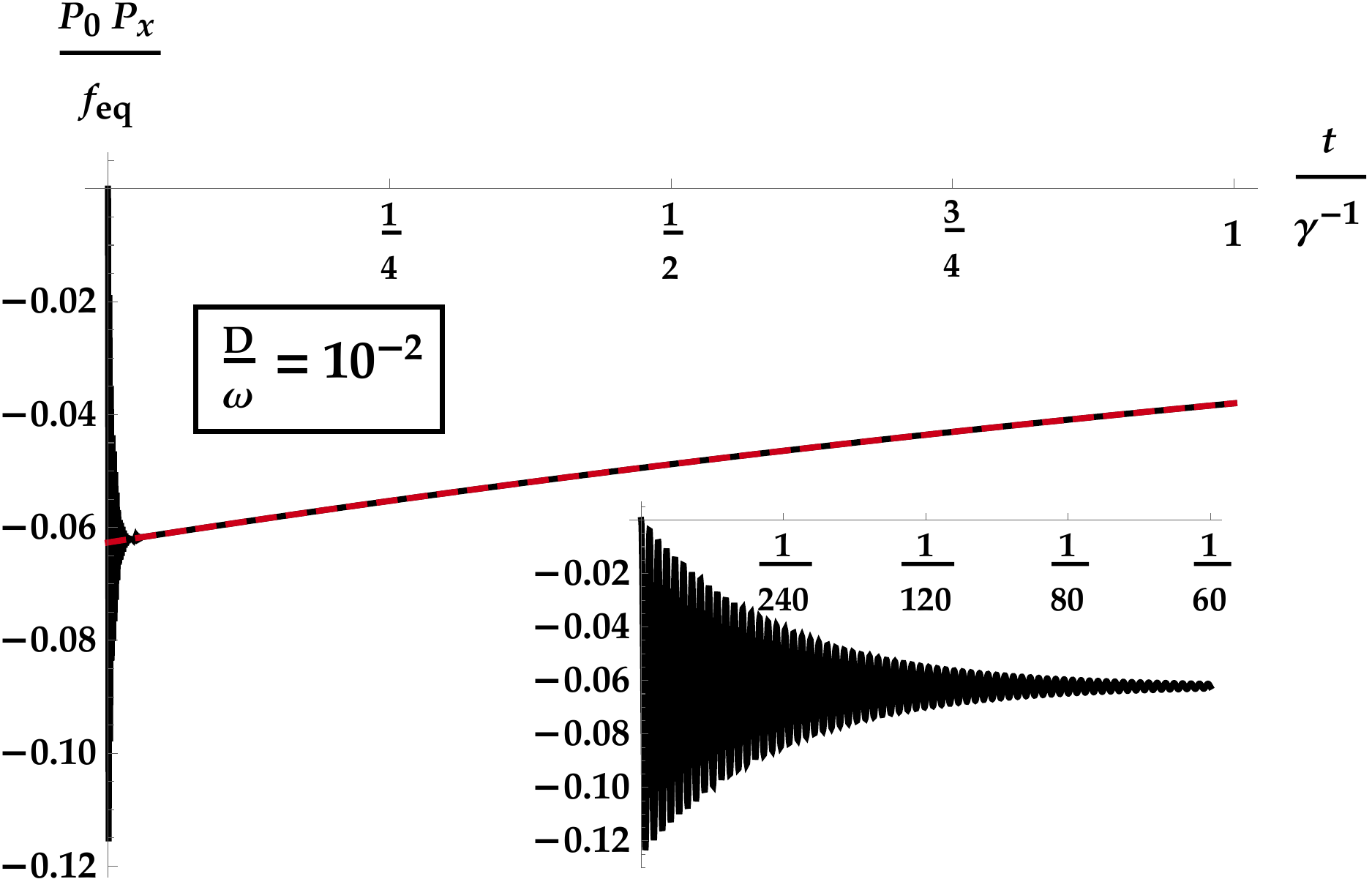}
}
\end{subfigure}
\begin{subfigure}{
\centering
\includegraphics[width=.47\textwidth]{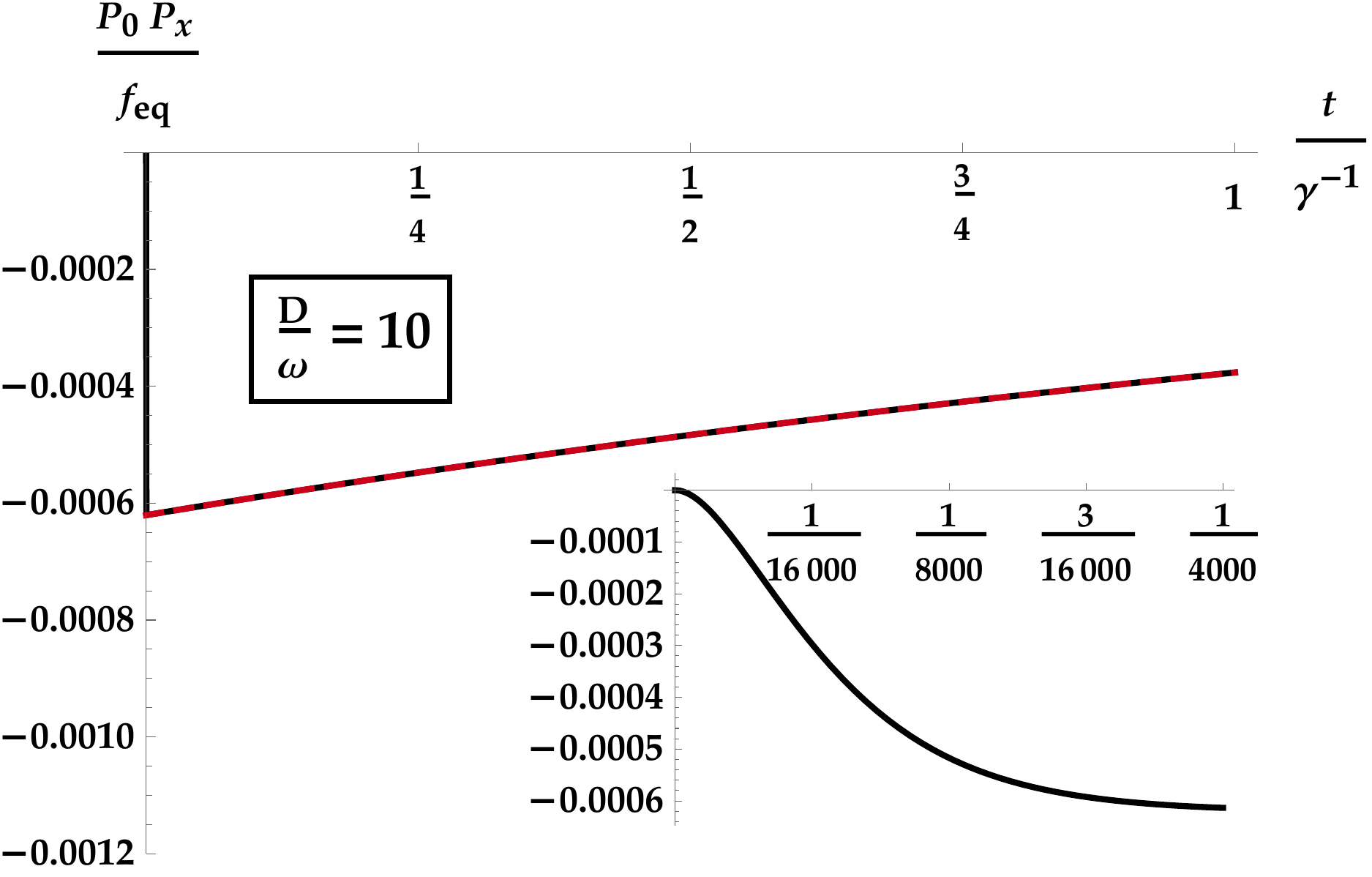}
}
\end{subfigure}
\caption{Comparison of the relaxation [Eq.~\eqref{relax}; dashed, red curve] and QKE [Eq.~\eqref{qke1}; black] solutions for $P_0 (t) P_x (t) = \rho_{as}(t) + \rho_{sa}(t)$. As in Fig.~\ref{magplot}, the onset time of the relaxation solution is fit by hand.}
\label{pxplots}
\end{figure*}

\section{Numerical analysis \label{numerics}}

In this section we numerically study the validity of the quantum relaxation-time approximation, including the sterile neutrino Boltzmann equation implied by it. For simplicity we begin by assuming time-independent mixing and scattering parameters $\omega$, $\sin^2 2\theta$, and $D$. We then go on to introduce a time-dependent potential and follow the system through resonance.

Fig.~\ref{fsplots} compares the solutions of the Boltzmann equation [Eq.~\eqref{boltz1}] and the QKEs [Eqs.~\eqref{poleom}] for two choices of the ratio $D/ \omega$. In both cases the mixing angle is taken to be $\theta = \pi / 100$, and in the Boltzmann equation $f_a$ is always set equal to $f_a^\textrm{eq}$. The insets show that the solutions are discrepant at very early times before the QKE solution settles into the decay mode [Eq.~\eqref{dfsdt}] on which the Boltzmann equation is based. Once it does so, both solutions grow linearly in time, as expected when $f_s \ll f_a^\textrm{eq}$. As production proceeds, the early-time discrepancy becomes less important as a fraction of the sterile abundance, and on times longer than $\gamma^{-1}$ the evolution as a function of $t' = \gamma t$ is virtually independent of the chosen parameters:
\begin{equation}
f_s (t') = f_a^\textrm{eq} \left( 1 - e^{- \frac{t'}{2}} \right).
\end{equation}

\begin{figure*}
\centering
\begin{subfigure}{
\centering
\includegraphics[width=.41\textwidth]{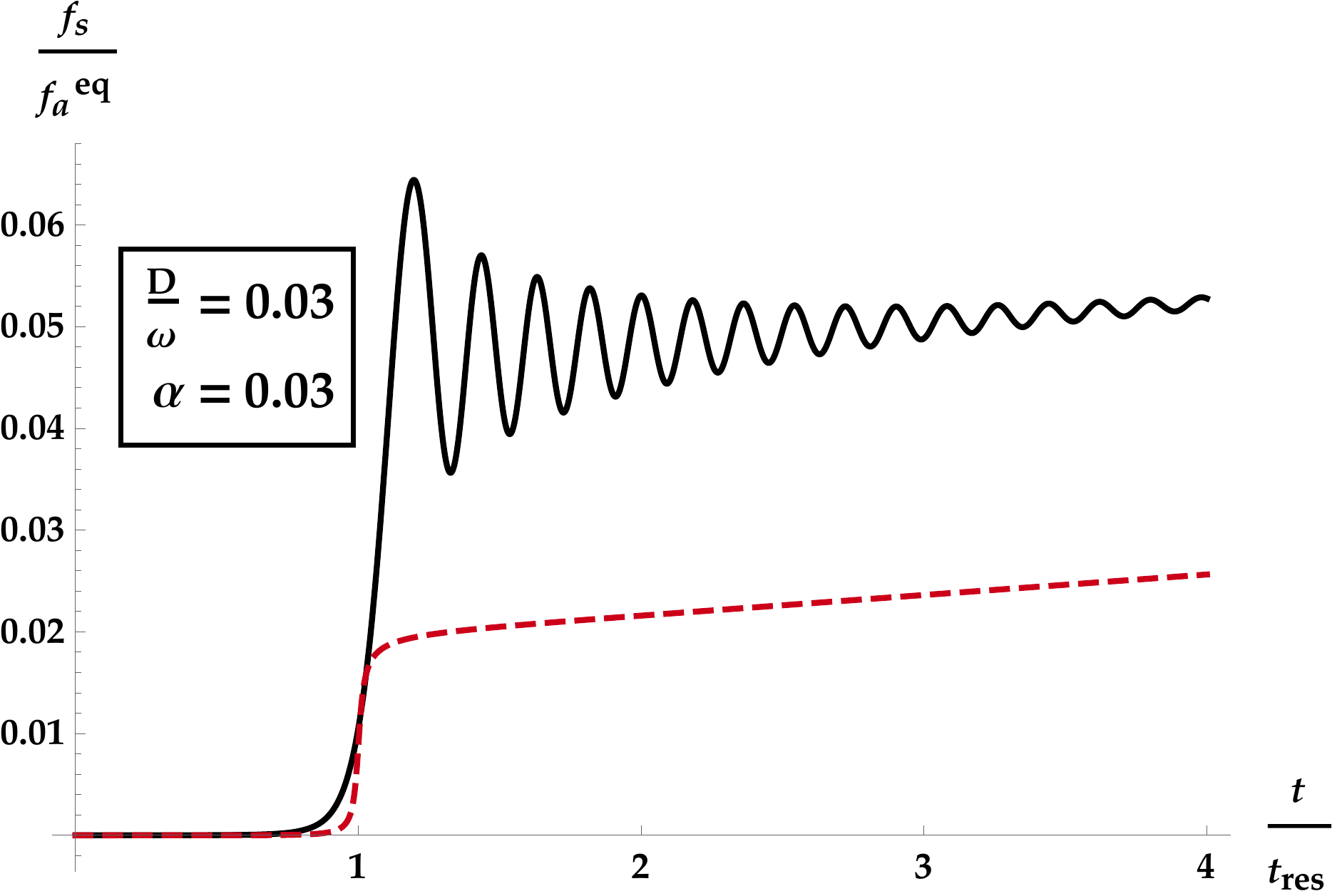}
}
\end{subfigure}
\begin{subfigure}{
\centering
\includegraphics[width=.41\textwidth]{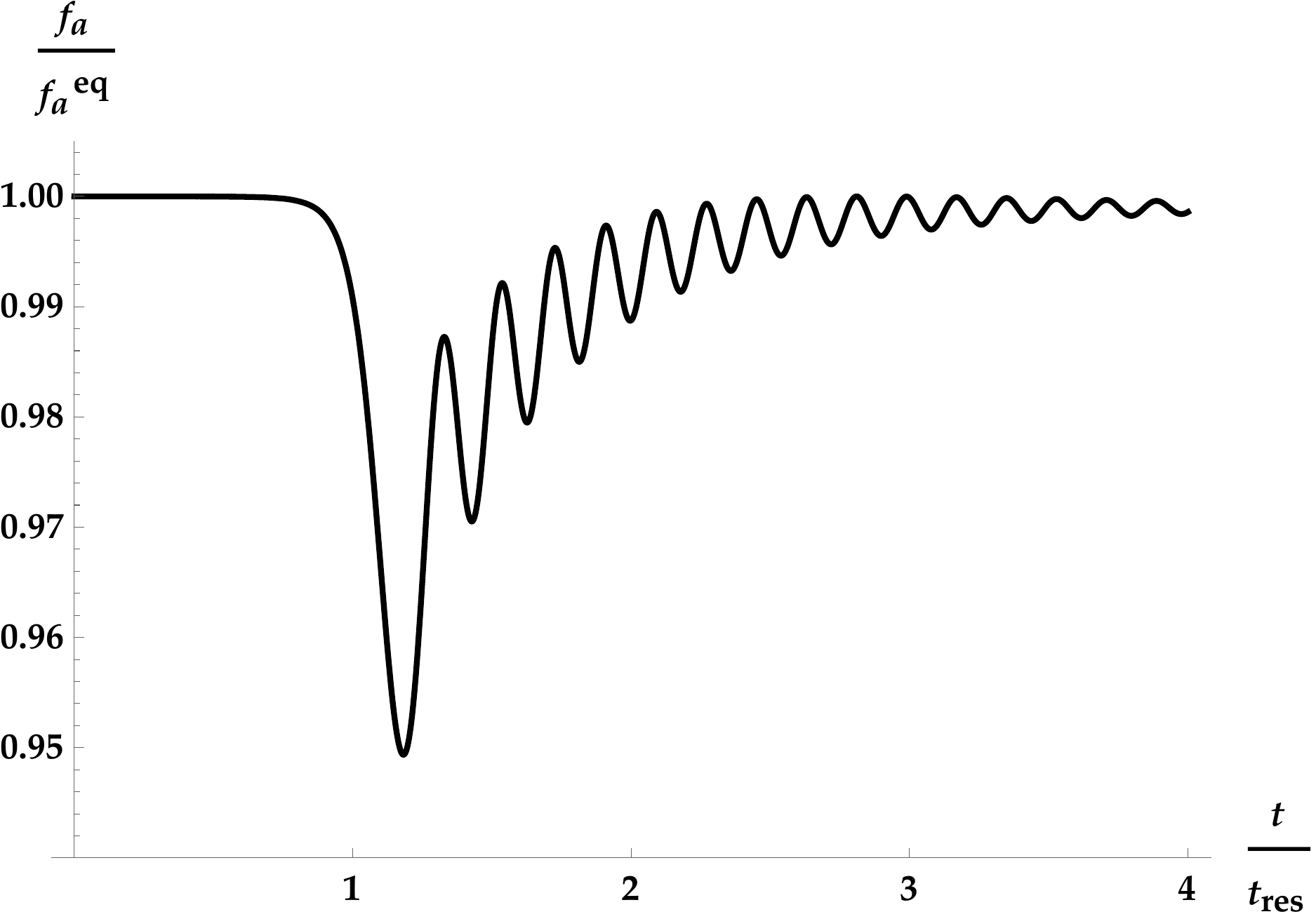}
}
\end{subfigure}

\begin{subfigure}{
\centering
\includegraphics[width=.41\textwidth]{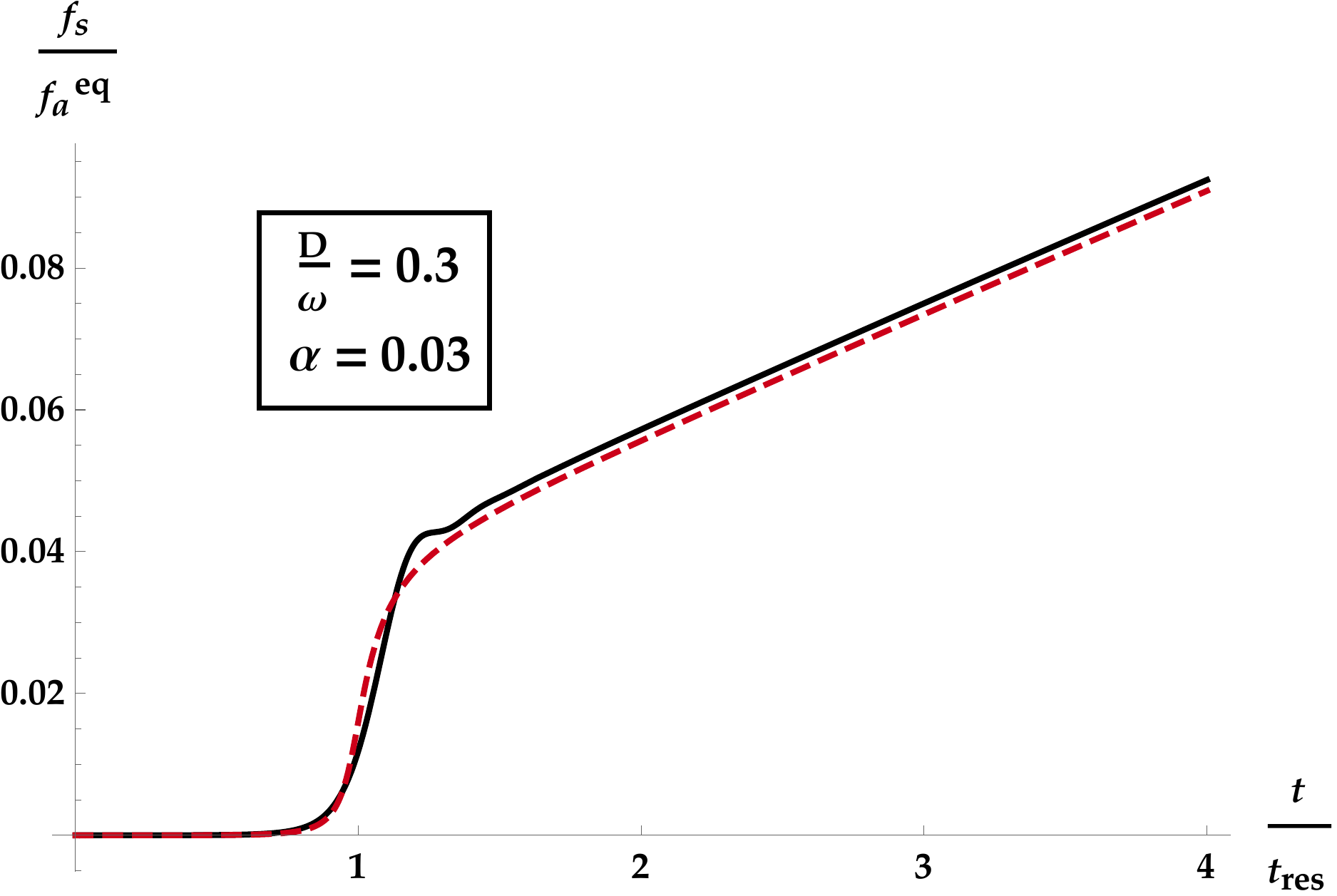}
}
\end{subfigure}
\begin{subfigure}{
\centering
\includegraphics[width=.41\textwidth]{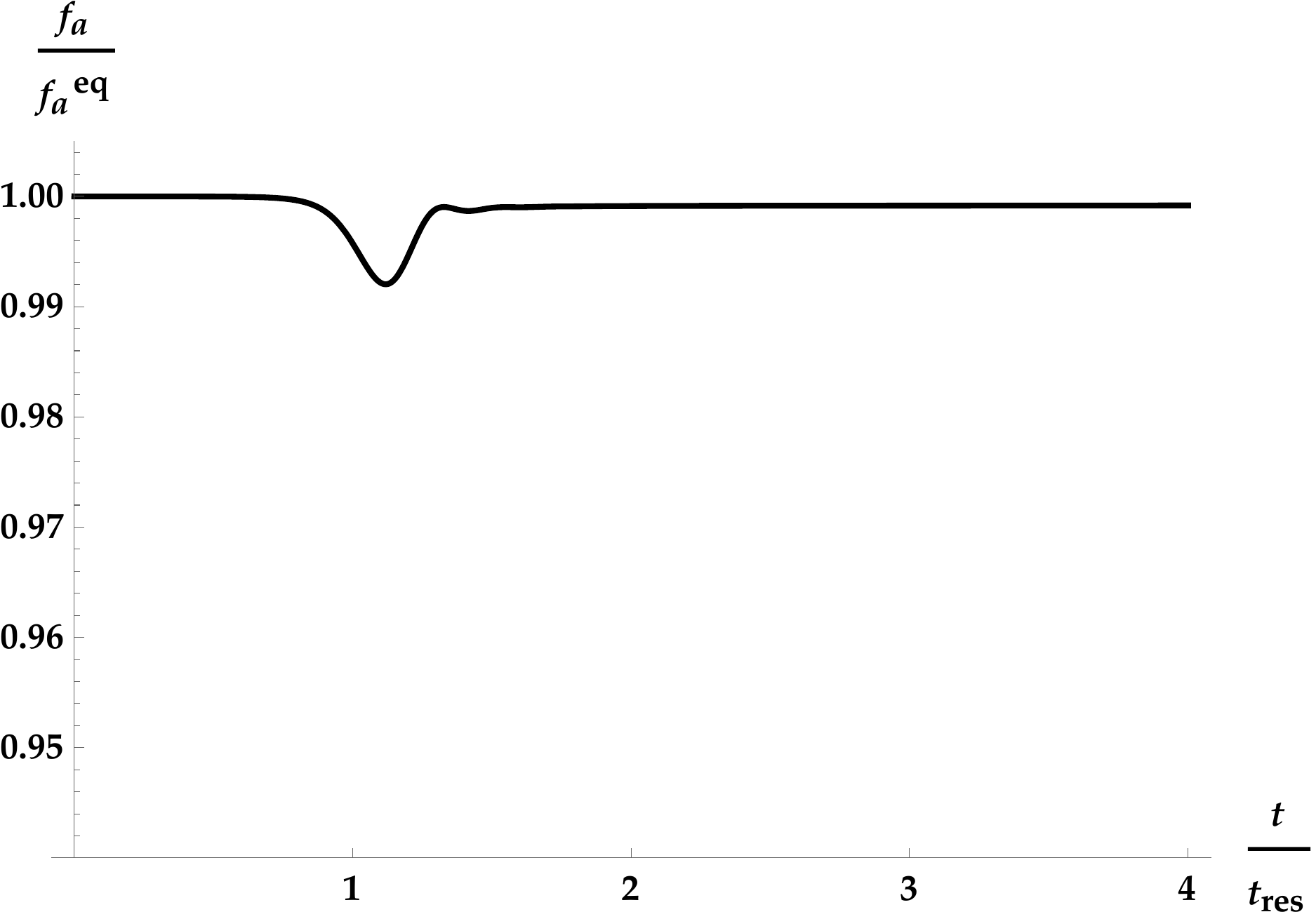}
}
\end{subfigure}

\begin{subfigure}{
\centering
\includegraphics[width=.41\textwidth]{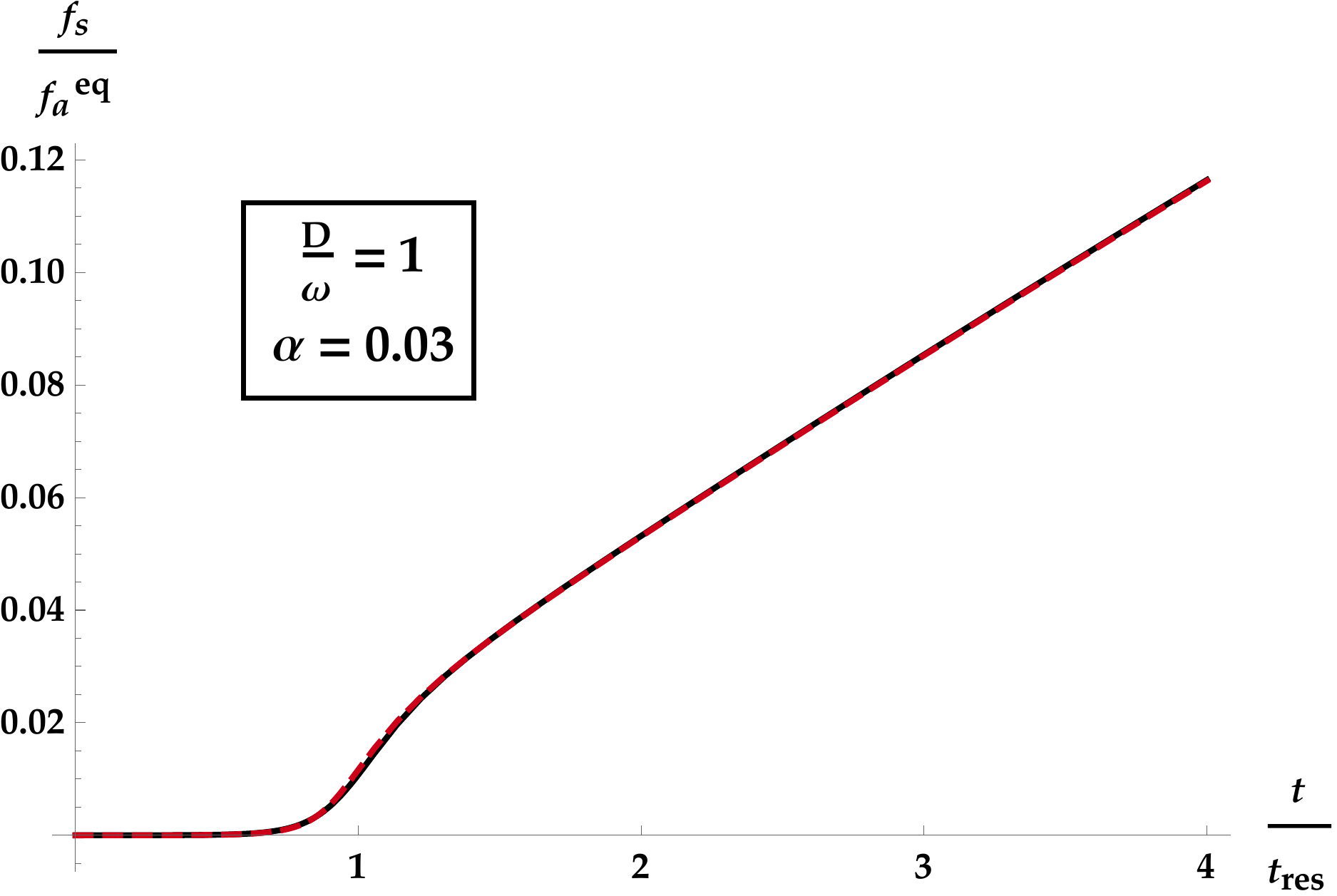}
}
\end{subfigure}
\begin{subfigure}{
\centering
\includegraphics[width=.41\textwidth]{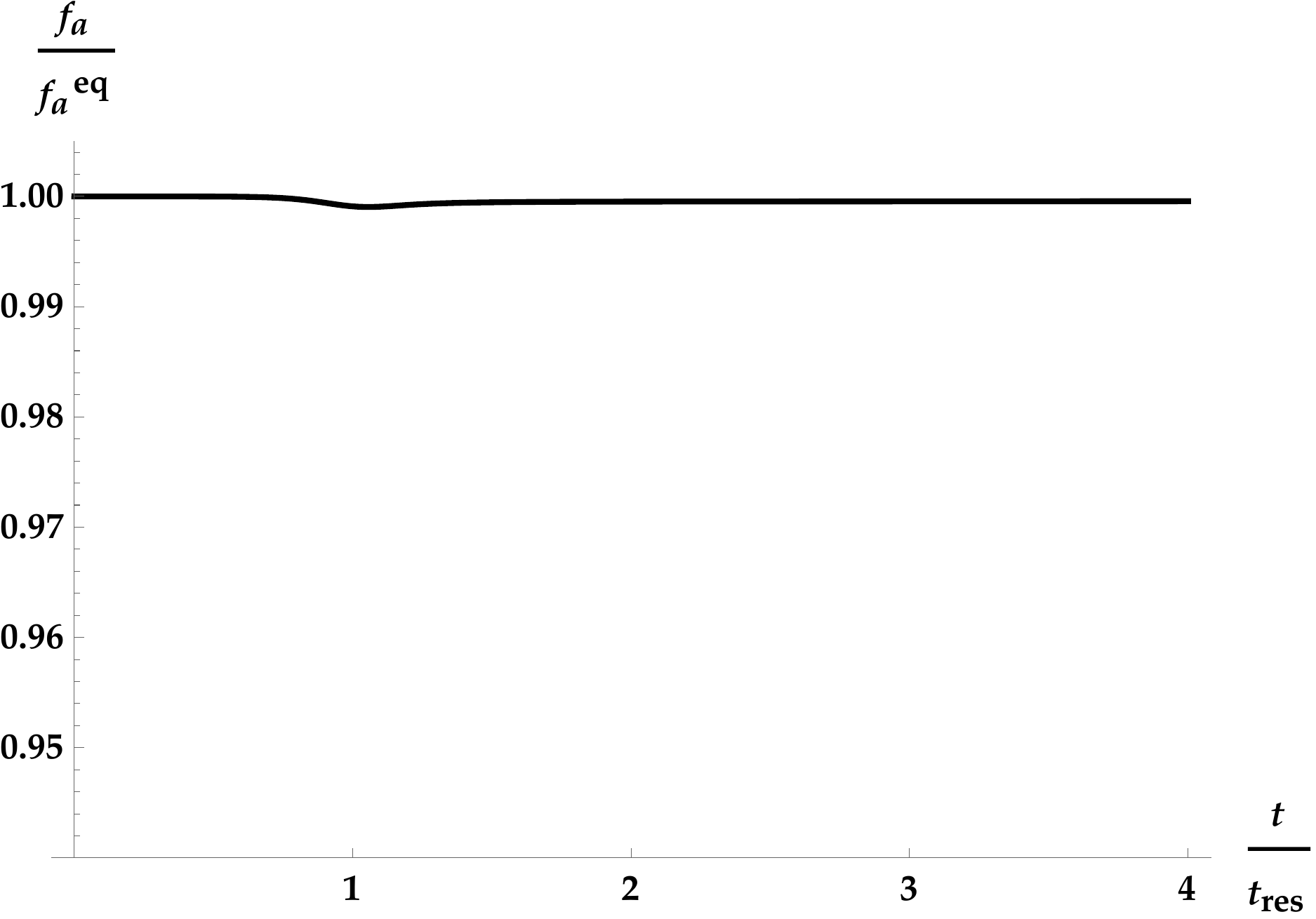}
}
\end{subfigure}

\begin{subfigure}{
\centering
\includegraphics[width=.41\textwidth]{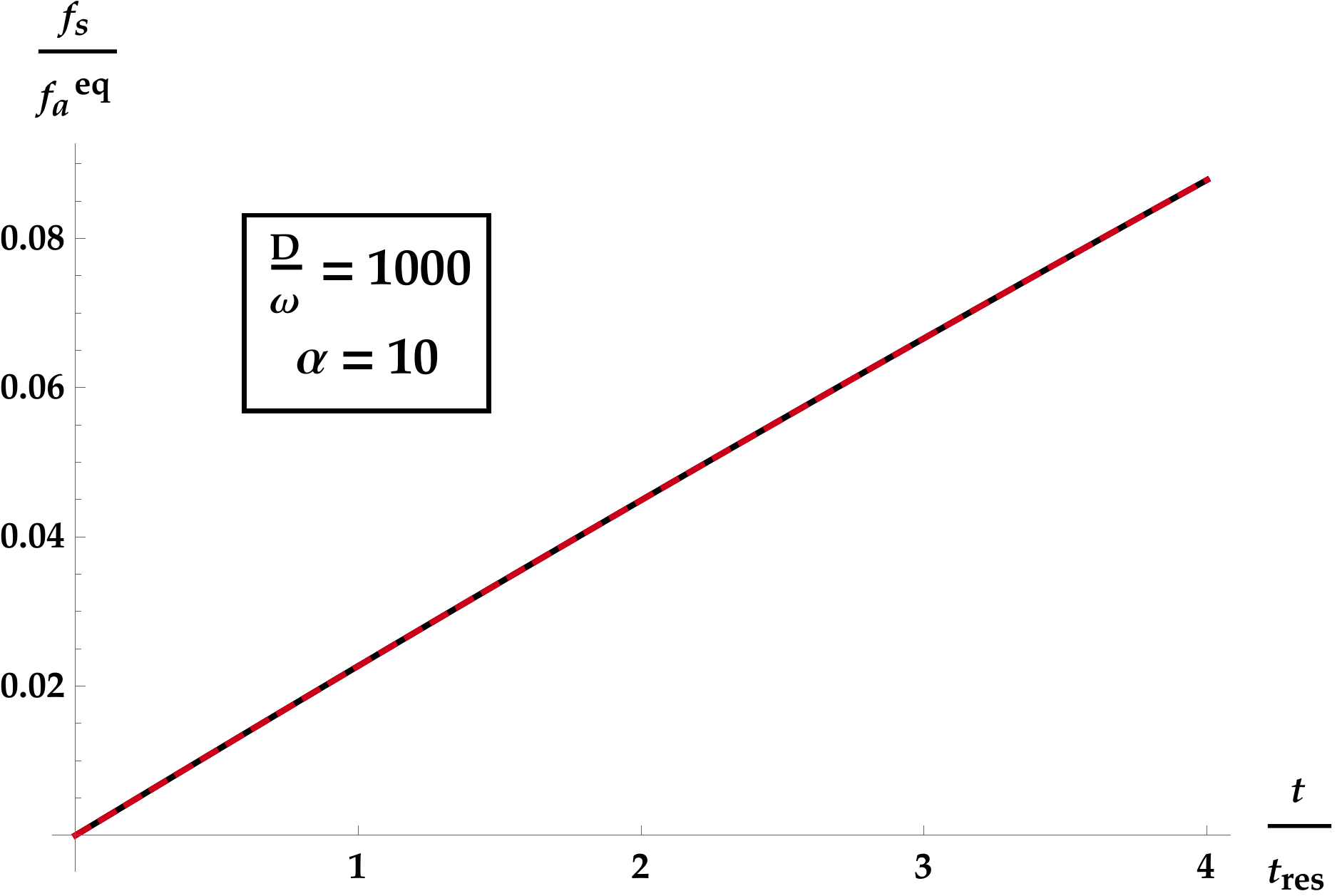}
}
\end{subfigure}
\begin{subfigure}{
\centering
\includegraphics[width=.41\textwidth]{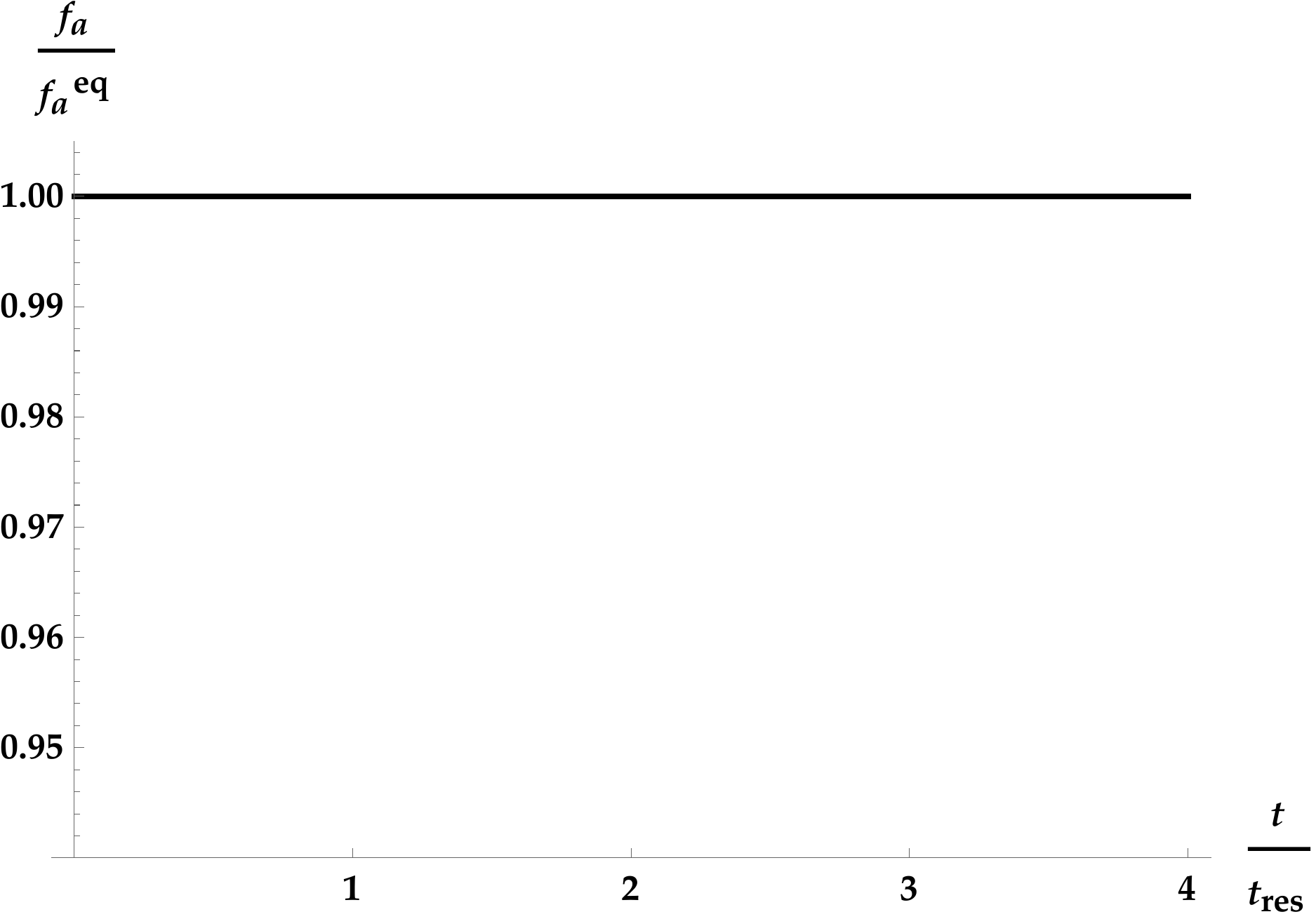}
}
\end{subfigure}
\caption{Left panels: Comparison of the Boltzmann and QKE solutions for $f_s (t)$ in the presence of a time-dependent potential $V(t) = V_0 e^{-\nu t}$. Right panels: QKE solution for $f_a (t)$; the Boltzmann solution (not shown) is simply $f_a (t) = f_a^\textrm{eq}$. Unlike in the previous figures, time is plotted in units of $t_\textrm{res}$, defined by $\omega \cos 2\theta = V(t_\textrm{res})$. From top to bottom, the coherence parameter $\kappa$ is $4.4$, $0.44$, $0.13$, and $4.0 \times 10^{-7}$. Agreement between the Boltzmann and QKE solutions is good if $\kappa \lesssim 1$. When that condition is satisfied, the active species interacts strongly enough with the medium to ensure that $\rho$ does not deviate too strongly from the relaxation solution.}
\label{fsfa}
\end{figure*}

For cosmological applications it is typically undesirable to have the density of sterile neutrinos approach the thermal value, and the insets in Fig.~\ref{fsplots} are therefore the relevant comparison. Achieving agreement on these shorter time scales requires $\theta$ to be small. Fig.~\ref{fslargeplots} illustrates this point: the same quantities are plotted here as in Fig.~\ref{fsplots}, but now with $\theta = \pi / 5$. Early-time discrepancies are greatly exacerbated.

The relaxation ansatz asserts that $f_a$ should decay exponentially toward $f_a^\textrm{eq}$. Since a small mixing angle inhibits $f_a$ from ever deviating greatly from the equilibrium value, the ansatz also implies a delicate near-cancellation between the growth of $P_0$ and the decay of $P = | \mathbf{P} |$. Fig.~\ref{magplot} verifies that both of these expectations are indeed borne out in the case $D / \omega = 10^{-2}$. The result is similar for stronger damping.

Interestingly, it was shown in Ref.~\cite{dolgov2002b} that the Boltzmann equation can be derived from the assumption that $\rho_{as}$ and $\rho_{sa}$ are both constant. Despite its expedience, that approximation is not an accurate description of how the active-sterile coherence develops, particularly on a $\gamma^{-1}$ time scale. Fig.~\ref{pxplots} shows that in fact the real part declines throughout production and is well fit by the relaxation ansatz. The imaginary part is similar.

Cosmological production of sterile neutrinos involves time-dependent parameters of course, and many scenarios of interest involve resonant mixing in particular \cite{shi1999, abazajian2001b, abazajian2014, kishimoto2008, venumadhav2016, johns2016, johns2018, johns2019}. It is well-known that the Boltzmann equation is inadequate when the system passes adiabatically and coherently through a resonance \cite{abazajian2001b, abazajian2005, kishimoto2008}. To illustrate this claim, and to make a connection with the foregoing analysis, we add a potential $V(t) \mathbf{z}$ to $\omega \mathbf{B}$, with
\begin{equation}
V (t) = V_0 e^{- \nu t}.
\end{equation}
The adiabaticity parameter \cite{abazajian2005} is defined to be
\begin{equation}
\alpha = \omega \mathcal{H} \sin 2\theta \tan 2\theta,
\end{equation}
where $\mathcal{H}$ is the potential scale height,
\begin{equation}
\mathcal{H} = \bigg| \frac{1}{V} \frac{dV}{dt} \bigg|^{-1} = \frac{1}{\nu}.
\end{equation}
We also define $\kappa = D^{-1} / \mathcal{H}$, or in terms of the adiabaticity,
\begin{equation}
\kappa = \frac{\omega}{D} \frac{\sin 2\theta \tan 2\theta}{\alpha}
\end{equation}
While $\alpha$---proportional to the number of \textit{oscillation lengths} that fit within a resonance width---sets the probability of a neutrino coherently transitioning between energy eigenstates at resonance, the coherence parameter $\kappa$---the number of \textit{mean free paths} that fit within a resonance width---indicates the degree to which scattering affects the evolution.

In Fig.~\ref{fsfa} we compare the Boltzmann and QKE solutions for several choices of $D / \omega$ and $\alpha$. Each panel on the right shows $f_a (t)$ calculated with the same parameters used in the panel to its left. The extent to which $f_a$ dips away from the equilibrium value at resonance indicates the inapplicability of the relaxation ansatz during this period of production. The top three rows show nonadiabatic resonances with increasing rates of decoherence. The bottom row shows an adiabatic resonance with a very large decoherence rate. Regardless of the adiabaticity, production through the resonance is flattened out as $D / \omega$ increases, and the accuracy of the Boltzmann solution improves. The key to the Boltzmann equation being a valid approximation is always that $\kappa$ be small, to ensure that the relaxation ansatz for $\rho$ is not too badly violated at resonance.

\section{Summary \label{conclusion}}

A few different derivations of the sterile neutrino Boltzmann equation can be found in the literature. Aside from its simplicity, the relaxation-time approach is notable in that it is based on an accurate description of the full quantum dynamics of active--sterile mixing. Replicating Eq.~\eqref{boltz1} is one implication, but the approximation describes equally well the repopulation of the active species and the decay of coherence.

We have numerically shown that the limitations of Eq.~\eqref{boltz1} reflect the assumptions used here to derive it: the Boltzmann solution works best when $\theta_m$ is small, and it comes up short when resonance coherently steers the system away from relaxation. Cosmological sterile-neutrino production is just one scenario of interest that is covered, at least partially, by this range of validity. A similar analysis can be applied to the mixing of active states in the presence of unequal chemical potentials---a circumstance that is realized in supernovae, binary neutron-star mergers, and possibly the early universe. In light of the computational obstacles faced in these environments, there is a clear need for good quantum-kinetic approximations.

\begin{acknowledgments}
The author gratefully acknowledges conversations with Daniel Blaschke, Vincenzo Cirigliano, George Fuller, Evan Grohs, Chad Kishimoto, Mark Paris, and Shashank Shalgar. This work was supported by NSF Grant No. PHY-1614864.
\end{acknowledgments}

\bibliography{all_papers}

\end{document}